\documentclass[twocolumn, aps, prl, amsmath, amssymb, a4paper, floatfix]{revtex4-2}

\usepackage{graphicx} 
\usepackage{xcolor}
\usepackage[utf8x]{inputenc}
\usepackage{physics}
\usepackage{siunitx}
\usepackage{textgreek}
\usepackage[version=4]{mhchem}
\usepackage{comment}
\usepackage{lipsum}
\usepackage{upgreek}
\usepackage[hidelinks]{hyperref}

\begin{document}
\title{Faraday Wave Singularities Trigger Microbubble Jetting}

\author{Marco Cattaneo}
\author{Louan Presse}
\author{Outi Supponen}

\affiliation{Institute of Fluid Dynamics, ETH Zürich, Zürich, Switzerland}

\begin{abstract}

Wall-attached bubbles can produce repeated jets under gentle ultrasound stimulation through the Faraday instability.
We identify three distinct jetting regimes defined by the jetting frequency and the bubble surface topology. 
We demonstrate that these jets form via flow-focusing singularities following two distinct collapse modes of the bubble interface: conical, producing a jet towards the substrate, or parabolic, generating a pair of oppositely directed jets.
Scaling laws governing these collapse events are derived, revealing a universal self-similar structure governed by inertia and capillarity.
Furthermore, we establish the dependence of the interface acceleration for jetting on driving frequency and characterise the jet speed as a function of Faraday wave height and bubble size.
These findings may inform the design of low-power biofilm removal ultrasound systems and contribute to improved safety in targeted drug delivery.
\end{abstract}

\maketitle

Bubbles undergoing rapid shape changes can generate high-speed liquid jets, posing hazards such as contaminant dispersion during the bursting of bubbles \cite{Veron2015OceanSpray} or material erosion in hydraulic systems experiencing cavitation \cite{Arndt1981CavitationStructures}.
However, when harnessed effectively, bubble jets benefit applications like surface cleaning \cite{Reuter2017MembraneBubbles}, chemical reaction enhancement \cite{Suslick1995ApplicationsChemistry}, biofilm removal \cite{Lattwein2020Sonobactericide:Infections}, targeted tissue ablation \cite{Xu2024Histotripsy:Ultrasound}, and drug delivery \cite{Shakya2024Ultrasound-responsiveDelivery}.
Bubble jets form via two mechanisms.
The first, inertial jetting, results from steep pressure gradients caused, for example, by strong acoustic drivings \cite{Rossello2018AcousticallyJets,Cleve2019ContrastUltrasound}, shock waves \cite{Ohl2003Shock-Wave-InducedBubbles,Bokman2023ScalingWave}, or asymmetric velocity distributions near boundaries \cite{Blake1987CAVITATIONBOUNDARIES.,Supponen2016ScalingBubbles}. 
The second, capillary-driven jetting, originates from flow-focusing singularities on the bubble interface during events like bubble bursting \cite{Duchemin2002JetSurface,Ghabache2014OnEjection}, bubble pinch-off \cite{Burton2005ScalingPinch-off,Gordillo2005AxisymmetricNumbers}, bubble coalescence \cite{Jiang2024AbyssCollisions}, or the collision of capillary waves on cavitation bubbles collapsing at very low stand-off distances from a wall \cite{Lechner2019FastStudy,Reuter2021SupersonicBubbles}.
In many practical applications, gas bubbles are driven by ultrasound, a non-invasive, cost-effective method well-suited for biomedical use.
Typically, jetting is initiated by strong ultrasound pressures, following the inertial jetting paradigm, which generates a transient, intense jet and subsequent bubble fragmentation.
A second jetting mechanism also exists, enabling jet formation at ultrasound pressures roughly one order of magnitude lower than those required for inertial jets.
Though early evidence exists \cite{Crum1979SurfaceBubbles,Prabowo2011SurfaceBubbles,Vos2011,Biasiori-Poulanges2023SynchrotronCavitation}, this phenomenon was only recently fully recognised \cite{Cattaneo2025CyclicDeliveryb}.
It arises from repeated bubble oscillations under ultrasound, triggering the Faraday instability and forming a standing wave pattern. 
The periodic collapse of pattern crests generates repeated jets without necessarily causing fragmentation. 
Since this jet formation does not stem from a pressure gradient, it falls under jets driven by capillary phenomena.
This second jetting mechanism can be considered as the spherical counterpart to the Faraday instability-induced jets observed on vertically vibrated flat liquid surfaces, first reported by Longuet-Higgins \cite{Longuet-Higgins1983BubblesSurface} and later studied by Lathrop’s group \cite{Goodridge1996ThresholdWaves,Goodridge1997ViscousWaves,Zeff2000SingularitySurface}.
However, a key difference lies in the driving mechanism of the Faraday waves: on flat surfaces, they result from external vibrations, whereas in bubbles, they stem from the bubble's own oscillatory motion, induced by ultrasound due to its compressibility.
Our previous works \cite{Cattaneo2025CyclicDeliveryb,Cattaneo2025ShapeBubblesb} demonstrated that Faraday wave-induced jets occur in both wall-bounded and freely oscillating bubbles.
We also characterised the standing wave patterns and their evolution, though the exact jet origin remained elusive due to rapid dynamics.
Here, using large wall-attached bubbles and ultra-high-speed imaging, we effectively slow down the process, expanding our characterisation of this jetting mechanism. 
This reveals its origin in flow-focusing singularities and links it to other capillary phenomena involving bubbles and jets.

\begin{figure}[t]
    \includegraphics[width = \columnwidth]{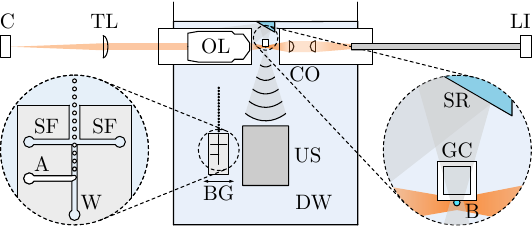}
     \caption{Experimental setup. (A) Air, (B) Bubble, (BG) Bubble generator, (C) Camera, (CO) Condenser, (DW) Deionised water, (GC) Glass capillary, (LI) Laser illuminator, (OL) Objective lens, (SF) Sheath flow, (SR) Sound reflector, (TL) Tube lens, (US) Ultrasound transducer, (W) Water.}
     \label{fig:Figure1}
\end{figure}
The experimental setup is sketched in Fig.~\ref{fig:Figure1}.
A monodisperse stream of air bubbles with equilibrium radii $R_0 = \SIrange{5}{80}{\micro\meter}$ is produced via a microfluidic T-junction chip (for details, see Ref.~\cite{Cattaneo2025ShapeBubblesb}) immersed in water, with density $\rho = \SI{998}{\kilo\gram\per\cubic\meter}$, dynamic viscosity $\mu = \SI{0.98}{\milli\pascal\second}$, surface tension $\sigma = \SI{72}{\milli\newton\per\meter}$, and temperature $T = \SI{22}{\celsius}$.
Individual bubbles are placed at the bottom of a squared borosilicate glass capillary, whose hollow structure reduces acoustic reflections compared to full substrates.
The bubble is driven acoustically at a frequency $f_{\rm d} = \SI{100}{\kilo\hertz}$ using an ultrasound transducer (PA2366, Precision Acoustics), and its response is captured from a side-view perspective.
The imaging system includes a 20-mm focal length objective (N10XW-PF, Nikon) and a 600-mm tube lens (TL600-A, Thorlabs), achieving $30\times$ magnification.
Recordings are performed with an ultra-high-speed camera (HPV-X2, Shimadzu) at 0.5–10 MHz.
Backlight illumination is provided by a pulsed diode laser (CAVILUX Smart UHS) and focused onto the sample using a custom condenser.
A glass reflector prevents sound reflections at the water-air interface.

\begin{figure}[t]
    \includegraphics[width = \columnwidth]{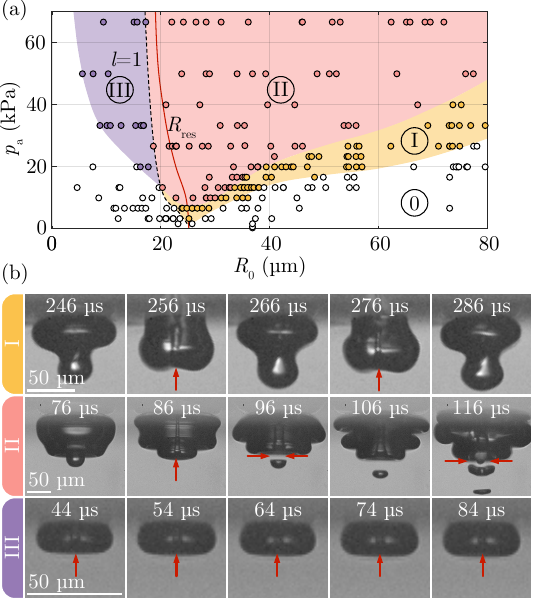}
     \caption{ 
     (a) Map of jetting regimes of wall-attached bubbles driven by ultrasound for a range of bubble equilibrium radii and driving pressures. 
     (0) Spherical oscillations without jetting.   
     (I) Half-harmonic jets directed towards the substrate.  
     (II) Half-harmonic bubble splitting with Worthington-like jet pairs formation.  
     (III) Harmonic jets directed against the wall.  
     The coloured area represents the experimentally-determined parameter space for bubble jetting.
     The red line indicates the resonant bubble size (for details, see Ref.~\cite{Cattaneo2025ShapeBubblesb}).  
     The dashed line marks the theoretical transition between bubble sizes exhibiting shape modes of degree $l=1$ and those with higher-degree modes (for details, see Ref.~\cite{Cattaneo2025ShapeBubblesb}).  
     (b) Image sequences of the three jetting regimes.  Red arrows highlight jets or bubble splitting.
     }\label{fig:Figure2}
\end{figure}

Figure \ref{fig:Figure2} provides an overview of jetting regimes for wall-attached bubbles across various equilibrium radii and ultrasound pressures.
Three distinct jetting regimes, along with the base regime, are identified:
(0) at low driving pressures, the bubble exhibits spherical oscillations.
These oscillations, in turn, harmonically modulate the bubble meniscus at the wall, producing weak, stationary waves on the bubble surface (Supplementary Material Video 1), as we previously reported in Ref.~\cite{Cattaneo2025ShapeBubblesb}.
No jetting occurs.
(I) At higher driving pressures, bubble oscillation destabilises the interface, a phenomenon known as Faraday instability, leading to half-harmonic standing wave patterns or shape modes. Above a critical amplitude, during each wave cycle, the bottom lobe of the shape mode folds in so vigorously that it generates a high-speed jet directed at the substrate [Fig.~\ref{fig:Figure2}(bI), Supplementary Material Video 2].
Since jetting is linked to the cyclicity of the shape mode, its frequency is half that of the ultrasound. Ultimately, non-axisymmetric shape modes may emerge, potentially halting jetting. We have extensively documented shape mode patterns in previous studies \cite{Cattaneo2025CyclicDeliveryb,Cattaneo2025ShapeBubblesb}. Notably, jets form at driving pressures as low as $p_{\rm a} \approx \SI{3}{\kilo\pascal}$ when bubble size matches the resonant size for the applied frequency.
\begin{figure*}[t]
    \includegraphics[width = \textwidth]{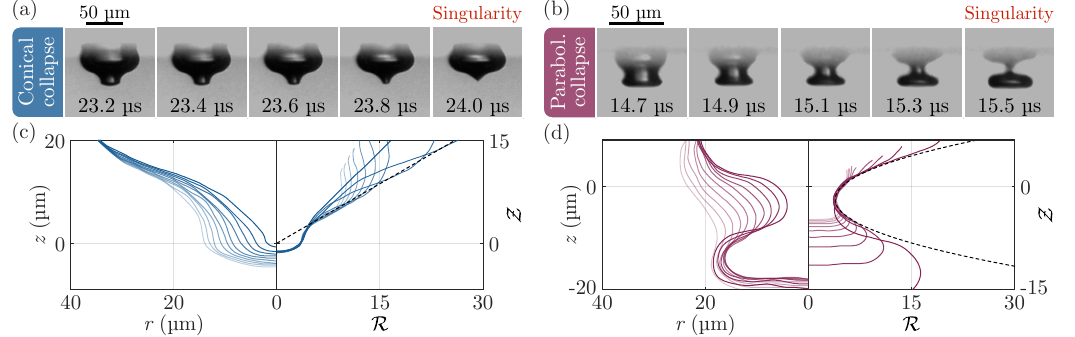}
     \caption{(a) Image sequence illustrating the conical collapse of the bottom lobe of the shape mode, culminating in a singularity from which a single jet emerges.
     (b) Image sequence illustrating the conical collapse of the shape mode bottom bowl,
     culminating in a singularity from which a jet pair emerges.
     (c),(d) Collapsing cavity profiles at ten time instants preceding the singularity, spaced \SI{0.1}{\micro\second} apart, extracted from (a),(b).
     Darker curves occur later in time.
     $r$ and $z$ are physical coordinates; 
     $\mathcal{R}$ and $ \mathcal{Z}$ are the rescaled ones. Dashed lines indicate conical and parabolic profiles.
     }
     \label{fig:Figure3}
\end{figure*}
(II) At even higher driving pressures, in the ultrasound cycle immediately following the first jet caused by bottom lobe folding, the surrounding bowl-shaped lobe also folds inward vigorously, splitting this bubble layer [Fig.~\ref{fig:Figure2}(bII), Supplementary Material Video 3].
At the point of closure, a pair of oppositely directed Worthington-like jets is ejected [in Fig.~\ref{fig:Figure2}(bII) these jets appear weak, mainly broadening the fluid column within the bubble, while Fig.~\ref{fig:Figure5}(b) highlights more pronounced jets].
This behaviour closely resembles the collapse of an air-filled cavity after a solid object impacts a water surface, also producing a pair of Worthington jets \cite{Gekle2009High-speedImpact,Gekle2010GenerationFormation}.
In the next shape mode cycle, a new bowl forms and collapses, splitting the bubble again. The bottom lobe ceases forming as the central liquid column, continuously sustained by the Worthington jets, inhibits its reformation.
(III) Small bubbles generate jets aimed at the substrate, cadenced with ultrasound frequency and thus termed harmonic [Fig.~\ref{fig:Figure2}(bIII), Supplementary Material Video 4]. 
Such bubbles do not exhibit standing wave patterns, preventing bubble splitting and preserving axisymmetry at high driving pressures, ensuring exceptional jetting stability.
By theoretically determining the shape modes wave number, or degree, as a function of bubble size and driving pressure (for details, see Ref.~\cite{Cattaneo2025ShapeBubblesb}), we find this regime to occur only when the bubble displays a shape mode of degree $l=1$ [Fig.~\ref{fig:Figure2}(a)].
Unlike the alternating-direction jetting observed in free bubbles exhibiting a $l=1$ mode \cite{Cattaneo2025CyclicDeliveryb}, the rigid substrate forces bubbles with this shape mode to consistently jet towards the substrate.
In general, compared to a freely oscillating bubble, the rigid substrate suppresses bubble oscillation at the contact surface.
As a result, Faraday waves develop mainly on the bubble side facing away from the substrate, directing jets predominantly towards it.

Figure~\ref{fig:Figure3}(a) shows the folding-in of the bottom lobe just before producing a jet towards the substrate. The collapsing bubble interface creates a singularity at its base, concentrating fluid kinetic energy along the central axis to drive jet formation. Similarly, Fig.~\ref{fig:Figure3}(b) depicts the folding-in of the bottom bowl just before generating a pair of opposing vertical jets. The interface collapse forms a singularity at the bowl centre, driving jet ejection.
To model these collapse modes, we assume that the bubble oscillatory motion is negligible on the interface  collapse timescale ($T_{\rm c}/T_{\rm o} < 0.1$, where $T_{\rm c}$ is the duration of the collapse event and $T_{\rm o}$ is the oscillation period), the bubble density is negligible relative to the surrounding liquid, and the fluid is incompressible and irrotational.
The fluid flow can then be described by the potential function $\phi(r,z,t)$, where $r$ and $z$ are the radial and axial coordinates, respectively.
The liquid is bounded by a free surface with height $h(r,t)$.
The behaviour of $\phi$ and $h$ is governed by the Laplace equation within the fluid, the kinematic equation for the fluid surface, and the Bernoulli equation applied at the free surface: 
\begin{equation}\label{eq:GoverningEquation}
\begin{cases}
\boldsymbol{\nabla}^{2}\phi = 0, \\[6pt]
\dfrac{\partial h}{\partial t} 
+ \dfrac{\partial h}{\partial r}\dfrac{\partial \phi}{\partial r} 
= \dfrac{\partial \phi}{\partial z}, 
\quad \text{on } z = h, \\[6pt]
\dfrac{\partial \phi}{\partial t} 
+ \dfrac{1}{2}\left|\boldsymbol{\nabla} \phi\right|^{2} 
+ \dfrac{\sigma}{\rho}\kappa = 0, 
\quad \text{on } z = h,
\end{cases}
\end{equation}
where $\kappa$ denotes the surface curvature.  
Following Keller and Miksis~\cite{Keller1983SurfaceFlows}, we non-dimensionalise the problem so that the formulation depends solely on dimensionless combinations of \(r\), \(z\), and the time-to-singularity \(\tau=t_0-t\).
With the ratio \( \sigma/\rho \) as the only physical parameter, dimensional analysis indicates that only two independent dimensionless variables can be constructed:
\begin{equation}\label{eq:VarScaling}
\mathcal{R} = (\rho / \sigma \tau^2 )^{1/3} r,   \quad  \mathcal{Z} = (\rho / \sigma \tau^2 )^{1/3} z.
\end{equation}
Accordingly, the potential and interface height admit the dimensionless forms
\begin{equation}\label{eq:FunScaling}
\begin{split}
\Phi(\mathcal{R}, \mathcal{Z}) = (\sigma^2\tau/\rho^2)^{-1/3}\phi(r,z,t), \\
\mathcal{H}(\mathcal{R}, \mathcal{Z})  = (\rho / \sigma \tau^2 )^{1/3} h(r,z,t) .
\end{split}
\end{equation}
Substituting Eqs.~\eqref{eq:VarScaling} and \eqref{eq:FunScaling} into Eq.~\eqref{eq:GoverningEquation} removes the time dependence, yielding a self-similar system for \(\Phi\) and \(\mathcal{H}\) in the variables \(\mathcal{R}\) and \(\mathcal{Z}\) (the full derivation of the self-similar scaling is provided in Supplemental Material \cite{SeeScaling.}).
For the first collapse mode, the interface height $h$ remains finite as $\tau \rightarrow 0$, implying $\mathcal{H} \sim \mathcal{R}$ and, consequently, that the self-similar interface is conical. 
Figure~\ref{fig:Figure3}(c) shows that the scaling found, proportional to $\tau^{2/3}$, indeed makes the bubble interface near the singularity self-similar, confirming the essential role of both inertia (first two terms in the Bernoulli equation) and capillarity (third term in the Bernoulli equation) in the observed dynamics.
Moreover, it demonstrates its approximately conical shape.
This geometry and inertio-capillary self-similar scaling are exactly the same of those observed when bubbles burst at a liquid free surface \cite{Duchemin2002JetSurface,Ghabache2014OnEjection,Lai2018BubbleProfiles} and in collapsing Faraday waves on a vibrating flat liquid bath \cite{Zeff2000SingularitySurface}, underscoring a common jet formation mechanism.
A similar flow-focusing singularity has also been observed in collapsing cavitation bubbles in extreme proximity to a wall, where it generates supersonic jets \cite{Lechner2019FastStudy,Reuter2021SupersonicBubbles,Bumann2023InvestigationSimulation,Sieber2023CavitationMaterials}.
By contrast, for the second collapse mode, the top and bottom cavity walls flatten as $\tau \rightarrow 0$, indicating that $\mathcal{H} \sim \mathcal{R}^n$, with $0<n<1$. 
Figure~\ref{fig:Figure3}(d) shows that this mode also follows an inertio-capillary scaling, producing a self-similar geometry near the singularity through most of the collapse.
The bubble interface adopts an approximately parabolic shape, aligning with the expected sublinear dependence on $\mathcal{R}$.
However, just before the singularity occurs, the axial component of the interface deviates from self-similarity and decreases at a slower rate.
This slower decay prevents the curvature of the parabolic profile from diverging, explaining why $h \nrightarrow 0$ as $\tau \rightarrow 0$.
This aligns with theoretical predictions on capillary pinch-off \cite{Leppinen2003CapillaryFluids}, where the self-similar solution near pinch-off becomes unstable when the outer fluid is much denser than the inner.
It is also consistent with experimental observations of air bubble pinch-off, where axial curvature grows more slowly than azimuthal curvature \cite{Thoroddsen2007ExperimentsPinch-off}.
While the parabolic profile of this second collapse mode resembles the pinching-off of air bubbles in water \cite{Burton2005ScalingPinch-off,Gordillo2005AxisymmetricNumbers}, the underlying dynamics differs: bubble pinch-off is driven solely by fluid inertia in low viscous fluids, resulting in the minimum radius scaling as $\tau^{1/2}$ (plus a logarithmic correction \cite{Gordillo2005AxisymmetricNumbers,Eggers2007TheoryCavity}), or dominated by viscosity in high viscous fluids, leading to a $\tau$ scaling.
By contrast, the parabolic collapse observed here arises from the Faraday instability, with an inherent inertio-capillary nature.
To our knowledge, this might be the first reported instance of bubble splitting exhibiting a $\tau^{2/3}$ scaling.
In both collapse modes, the bubble surface evolution after the singularity—when a jet forms—cannot be tracked due to light refraction causing black bands.
We note that, for Weber numbers $\text{We}\gg1$, the inertio-capillary balance may no longer hold and the dynamics become predominantly inertial.
This accelerates the interface collapse, approaching a $\tau^{1/2}$ scaling, particularly when bubble entrapment occurs, as demonstrated by Gordillo and Blanco–Rodríguez in the context of bubble bursting at free interfaces \cite{Gordillo2023TheoryJets}.

\begin{figure}[t]
    \includegraphics[width = \columnwidth]{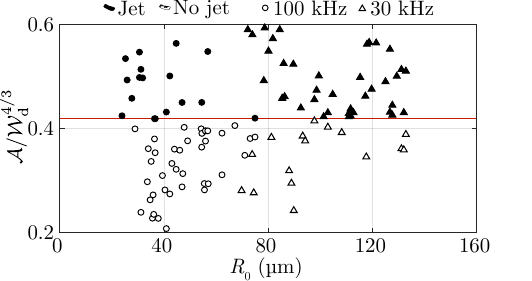}
     \caption{
     Bottom lobe acceleration for visually identified jetting events across bubble equilibrium radii at two ultrasound frequencies: \SI{30}{\kilo\hertz} (from our previous study \cite{Cattaneo2025ShapeBubblesb}) and \SI{100}{\kilo\hertz}.
     The dimensionless acceleration $\mathcal{A}$ is rescaled by $\mathcal{W}_{\mathrm{d}}^{4/3}$, with $\mathcal{W}_{\mathrm{d}}$ as the dimensionless driving angular frequency.
     The red line indicates the frequency-independent, acceleration threshold for jetting.
     }
     \label{fig:Figure4}
\end{figure}
To determine the acceleration of the bottom lobe at the onset of jetting and its dependence on driving frequency, we non-dimensionalise the interface acceleration \( a \), using the fluid properties \( \sigma/\rho \) and kinetic viscosity \( \nu \), following Goodridge \textit{et al.} for droplet-ejecting capillary waves \cite{Goodridge1996ThresholdWaves,Goodridge1997ViscousWaves}.
Only one independent dimensionless group can be formed:  
\begin{equation}
\mathcal{A} = a \nu^4 /(\sigma / \rho)^{3},
\end{equation}
which corresponds to the Morton number.
The interfacial acceleration is estimated from the lobe height $h_0$ as $a \approx h_0 \omega^2$, where $\omega=\omega_{\rm d}/2$ is the angular frequency of the Faraday wave and $\omega_{\rm d}$ is the angular driving frequency. 
The minimum wave height for jetting is proportional to the wavelength, \( h_0 \sim \lambda \) \cite{Michell1893Water,Schwartz1979NumericalWaves}.
Using the capillary-wave dispersion relation $\omega^2 = (\sigma/\rho)(2\pi/\lambda)^{3}$, it follows that the acceleration scales as $a \sim (\sigma/\rho)^{1/3}\omega_{\rm d}^{4/3}$.
Expressing this result in dimensionless form, we obtain
\begin{equation}
\mathcal{A} \sim \mathcal{W}_{\rm d}^{4/3},
\end{equation}
where $\mathcal{W}_{\rm d} = \omega_{\rm d}\nu^3/(\sigma/\rho)^{2}$ is the dimensionless angular driving frequency.
Figure~\ref{fig:Figure4} shows the measured dimensionless acceleration, \( \mathcal{A} \), rescaled by \( \mathcal{W}_{\rm d}^{4/3} \), for visually identified jetting events across various bubble radii and two ultrasound frequencies, \SI{30}{\kilo\hertz} (data from our previous study \cite{Cattaneo2025ShapeBubblesb}) and \SI{100}{\kilo\hertz}. 
The acceleration threshold for jetting collapses onto a single, frequency-independent value of approximately $0.4$, confirming the scaling found.
This behaviour holds in low-viscosity fluids, where surface tension dominates. 
By contrast, for higher-viscosity fluids (\( \mathcal{W}_{\rm d} \gg 10^{-5} \)), we anticipate a different scaling, \( \mathcal{A} \sim \mathcal{W}_{\rm d}^{3/2} \), based on prior studies of Faraday instability-induced jets on vertically vibrated liquid surfaces \cite{Hogrefe1998Power-lawWaves}.

\begin{figure}[t]
    \includegraphics[width = \columnwidth]{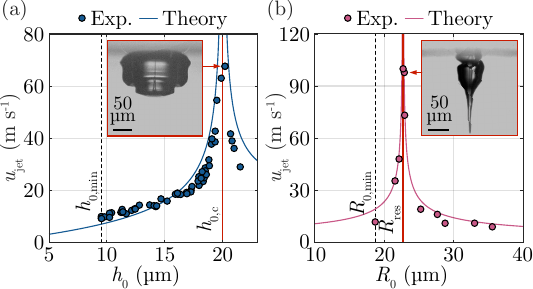}
     \caption{
     (a) Jet speed from conical collapse as a function of bottom lobe height. 
     (b) Jet pair speed from parabolic collapse as a function of bubble equilibrium radius at constant ultrasound pressure $p_{\rm a} = \SI{26.67}{\kilo\pascal}$.
     Dashed lines indicate the minimum threshold for jet formation. Red lines mark critical or resonant conditions.
     \label{fig:Figure5}
     }
\end{figure}
Figure~\ref{fig:Figure5}(a) depicts the speed of the jet emerging from a conical collapse as a function of the height $h_0$ of the bottom lobe of the shape mode.
The jet speed increases with wave height up to a critical value, $h_{\rm 0, c}\approx \SI{20}{\micro\meter}$, beyond which the bottom lobe pinches off a daughter bubble during its conical collapse.
As the lobe height exceeds this critical value, the extent of pinch-off increases, drawing more kinetic energy from the collapsing lobe and consequently reducing the jet speed.
Near $h_{\rm 0, c}$, jets reach speeds exceeding $u_{\rm jet} = \SI{60}{\meter\per\second}$. 
Below a minimum height $h_{\rm 0, min} \approx \SI{9.5}{\micro\meter}$, corresponding to the threshold acceleration identified in Fig.~\ref{fig:Figure4}, jets do not form.
Since the jetting dynamics is governed by the interplay between inertial and capillary forces, as well as wave breaking, the jet speed can be assumed to depend only on the Weber number $\text{We}$ and on the height deviation $(h_0-h_{\rm 0,c})$.
Accordingly, the jet speed can be expressed as
\begin{equation}
u_{\rm jet} \approx  \left( (h_0-h_{\rm 0, c})^{-1}\, \frac{\sigma}{\rho}\,\textrm{We}\right)^{1/2},
\end{equation}
where the Weber number is defined from the characteristics of the last smooth wave as  $\textrm{We}=\rho v^2 l/\sigma$, with $v = \omega h_0$ and $l=\lambda$.
This theoretical scaling for the jet speed shows good agreement with experimental observations and aligns well with the findings of Zeff \textit{et al.} for collapsing Faraday waves on a flat liquid interface \cite{Zeff2000SingularitySurface}.
Figure \ref{fig:Figure5}(b) depicts the speed of the jet pair emerging from a parabolic collapse as a function of the bubble equilibrium radius for a constant ultrasound pressure $p_{\rm a} = \SI{26.67}{\kilo\pascal}$.
The jet speed reaches its maximum at the resonant bubble radius, $R_{\rm res}$, where it can exceed $u_{\rm jet} = \SI{90}{\meter\per\second}$.
All tested bubbles exhibit an $l=2$ mode, adopting a purely oblate shape prior to undergoing parabolic collapse.
Consequently, the jetting dynamics is dictated by the bubble collapse rather than by wave dynamics, leading now to a jet speed that depends on the deviation of the bubble radius from its resonant value $(R_0 - R_{\rm res})$:
\begin{equation} 
u_{\rm jet} \approx \left( (R_0 - R_{\rm res})^{-1}\, \frac{\sigma}{\rho}\,\textrm{We}\right)^{1/2}.
\end{equation}
Here, the characteristic speed used to define the Weber number reflects the collapse-driving dynamics and is obtained from the ultrasound pressure as $v = (p_{\textrm a}/\rho)^{1/2}$, following standard practice in cavitation and acoustic bubble dynamics.
Notably, the jet directed away from the substrate can travel up to ten times the bubble equilibrium radius.
Jetting does not occur for bubble sizes below a threshold radius, $R_{\rm 0, min}$, marking the transition between shape mode orders $l = 1$ and $l = 2$. At this scale, the bubble is too small relative to the capillary wavelength $\lambda$ to allow a curvature sign change at the interface, preventing a parabolic collapse.
Future work could investigate how fluid viscosity influences jet speed across the two collapse regimes.
Prior studies of bubble bursting suggest that viscosity strongly modulates jet speed: moderate viscosities enhance wave focusing by damping short-wavelength perturbations, producing thinner, faster jets, while high viscosities slow the jet through increased shear and dissipation \cite{Deike2018DynamicsBubbles}.

In summary, we revealed that Faraday-instability-driven jetting 
originates from interfacial singularities emerging during the collapse of Faraday waves.
These singularities excel at concentrating kinetic energy from the surrounding fluid and converting it into a jet.
For instance, collapsing cavitation bubbles are known to produce inertial jets exceeding $\SI{100}{\meter\per\second}$. 
However, when the bubble is within one-tenth of its radius from a wall, a flow-focusing singularity emerges, producing jets at astonishing speeds over $\SI{1000}{\meter\per\second}$ \cite{Lechner2019FastStudy,Reuter2021SupersonicBubbles,Bumann2023InvestigationSimulation,Sieber2023CavitationMaterials}.
Here, instead, we harness this energy-focusing capability not to produce faster jets but to enable jetting under extremely mild acoustic driving.
In this regard, the Faraday instability offers a more efficient jet formation mechanism than inertial jetting.
Despite the lower driving, these jets still reach speeds of tens of meters per second, enough to damage hard materials. 
Moreover, the Faraday instability's gentle dynamics enables repeated jet formation. 
The conical and parabolic collapse geometries direct jets both towards and away from the substrate, enabling more versatile targeting than inertial jetting, including structures opposite the substrate.
Notably, Faraday-instability-induced jetting is the first known interfacial phenomenon where both of these collapse modes can occur. 
In conclusion, Faraday instability on wall-attached bubbles exhibits rich dynamics and promises practical applications, including low-power biofilm removal, enabling handheld devices, and safer targeted drug delivery.

\emph{Acknowledgments}---We thank ETH Zurich for financial support.

\emph{Data availability}---The data that support the findings of this article are openly available \cite{Cattaneo2025SourceHttps://doi.org/10.5281/zenodo.15085879}.

\bibliographystyle{apsrev4-2}
\bibliography{main/references.bib}

\begin{thebibliography}{46}%
\makeatletter
\providecommand \@ifxundefined [1]{%
 \@ifx{#1\undefined}
}%
\providecommand \@ifnum [1]{%
 \ifnum #1\expandafter \@firstoftwo
 \else \expandafter \@secondoftwo
 \fi
}%
\providecommand \@ifx [1]{%
 \ifx #1\expandafter \@firstoftwo
 \else \expandafter \@secondoftwo
 \fi
}%
\providecommand \natexlab [1]{#1}%
\providecommand \enquote  [1]{``#1''}%
\providecommand \bibnamefont  [1]{#1}%
\providecommand \bibfnamefont [1]{#1}%
\providecommand \citenamefont [1]{#1}%
\providecommand \href@noop [0]{\@secondoftwo}%
\providecommand \href [0]{\begingroup \@sanitize@url \@href}%
\providecommand \@href[1]{\@@startlink{#1}\@@href}%
\providecommand \@@href[1]{\endgroup#1\@@endlink}%
\providecommand \@sanitize@url [0]{\catcode `\\12\catcode `\$12\catcode `\&12\catcode `\#12\catcode `\^12\catcode `\_12\catcode `\%12\relax}%
\providecommand \@@startlink[1]{}%
\providecommand \@@endlink[0]{}%
\providecommand \url  [0]{\begingroup\@sanitize@url \@url }%
\providecommand \@url [1]{\endgroup\@href {#1}{\urlprefix }}%
\providecommand \urlprefix  [0]{URL }%
\providecommand \Eprint [0]{\href }%
\providecommand \doibase [0]{https://doi.org/}%
\providecommand \selectlanguage [0]{\@gobble}%
\providecommand \bibinfo  [0]{\@secondoftwo}%
\providecommand \bibfield  [0]{\@secondoftwo}%
\providecommand \translation [1]{[#1]}%
\providecommand \BibitemOpen [0]{}%
\providecommand \bibitemStop [0]{}%
\providecommand \bibitemNoStop [0]{.\EOS\space}%
\providecommand \EOS [0]{\spacefactor3000\relax}%
\providecommand \BibitemShut  [1]{\csname bibitem#1\endcsname}%
\let\auto@bib@innerbib\@empty
\bibitem [{\citenamefont {Veron}(2015)}]{Veron2015OceanSpray}%
  \BibitemOpen
  \bibfield  {author} {\bibinfo {author} {\bibfnamefont {F.}~\bibnamefont {Veron}},\ }\href@noop {} {\bibfield  {journal} {\bibinfo  {journal} {Annual Review of Fluid Mechanics}\ }\textbf {\bibinfo {volume} {47}} (\bibinfo {year} {2015})}\BibitemShut {NoStop}%
\bibitem [{\citenamefont {Arndt}(1981)}]{Arndt1981CavitationStructures}%
  \BibitemOpen
  \bibfield  {author} {\bibinfo {author} {\bibfnamefont {R.~E.}\ \bibnamefont {Arndt}},\ }\href@noop {} {\bibfield  {journal} {\bibinfo  {journal} {Annual Review of Fluid Mechanics}\ }\textbf {\bibinfo {volume} {13}} (\bibinfo {year} {1981})}\BibitemShut {NoStop}%
\bibitem [{\citenamefont {Reuter}\ \emph {et~al.}(2017)\citenamefont {Reuter}, \citenamefont {Lauterborn}, \citenamefont {Mettin},\ and\ \citenamefont {Lauterborn}}]{Reuter2017MembraneBubbles}%
  \BibitemOpen
  \bibfield  {author} {\bibinfo {author} {\bibfnamefont {F.}~\bibnamefont {Reuter}}, \bibinfo {author} {\bibfnamefont {S.}~\bibnamefont {Lauterborn}}, \bibinfo {author} {\bibfnamefont {R.}~\bibnamefont {Mettin}},\ and\ \bibinfo {author} {\bibfnamefont {W.}~\bibnamefont {Lauterborn}},\ }\href@noop {} {\bibfield  {journal} {\bibinfo  {journal} {Ultrasonics Sonochemistry}\ }\textbf {\bibinfo {volume} {37}} (\bibinfo {year} {2017})}\BibitemShut {NoStop}%
\bibitem [{\citenamefont {Suslick}(1995)}]{Suslick1995ApplicationsChemistry}%
  \BibitemOpen
  \bibfield  {author} {\bibinfo {author} {\bibfnamefont {K.~S.}\ \bibnamefont {Suslick}},\ }\href@noop {} {\bibfield  {journal} {\bibinfo  {journal} {MRS Bulletin}\ }\textbf {\bibinfo {volume} {20}} (\bibinfo {year} {1995})}\BibitemShut {NoStop}%
\bibitem [{\citenamefont {Lattwein}\ \emph {et~al.}(2020)\citenamefont {Lattwein}, \citenamefont {Shekhar}, \citenamefont {Kouijzer}, \citenamefont {van Wamel}, \citenamefont {Holland},\ and\ \citenamefont {Kooiman}}]{Lattwein2020Sonobactericide:Infections}%
  \BibitemOpen
  \bibfield  {author} {\bibinfo {author} {\bibfnamefont {K.~R.}\ \bibnamefont {Lattwein}}, \bibinfo {author} {\bibfnamefont {H.}~\bibnamefont {Shekhar}}, \bibinfo {author} {\bibfnamefont {J.~J.}\ \bibnamefont {Kouijzer}}, \bibinfo {author} {\bibfnamefont {W.~J.}\ \bibnamefont {van Wamel}}, \bibinfo {author} {\bibfnamefont {C.~K.}\ \bibnamefont {Holland}},\ and\ \bibinfo {author} {\bibfnamefont {K.}~\bibnamefont {Kooiman}},\ }\href@noop {} {\bibfield  {journal} {\bibinfo  {journal} {Ultrasound in Medicine and Biology}\ }\textbf {\bibinfo {volume} {46}} (\bibinfo {year} {2020})}\BibitemShut {NoStop}%
\bibitem [{\citenamefont {Xu}\ \emph {et~al.}(2024)\citenamefont {Xu}, \citenamefont {Khokhlova}, \citenamefont {Cho},\ and\ \citenamefont {Khokhlova}}]{Xu2024Histotripsy:Ultrasound}%
  \BibitemOpen
  \bibfield  {author} {\bibinfo {author} {\bibfnamefont {Z.}~\bibnamefont {Xu}}, \bibinfo {author} {\bibfnamefont {T.~D.}\ \bibnamefont {Khokhlova}}, \bibinfo {author} {\bibfnamefont {C.~S.}\ \bibnamefont {Cho}},\ and\ \bibinfo {author} {\bibfnamefont {V.~A.}\ \bibnamefont {Khokhlova}},\ }\href@noop {} {\bibfield  {journal} {\bibinfo  {journal} {Annual Review of Biomedical Engineering}\ }\textbf {\bibinfo {volume} {26}} (\bibinfo {year} {2024})}\BibitemShut {NoStop}%
\bibitem [{\citenamefont {Shakya}\ \emph {et~al.}(2024)\citenamefont {Shakya}, \citenamefont {Cattaneo}, \citenamefont {Guerriero}, \citenamefont {Prasanna}, \citenamefont {Fiorini},\ and\ \citenamefont {Supponen}}]{Shakya2024Ultrasound-responsiveDelivery}%
  \BibitemOpen
  \bibfield  {author} {\bibinfo {author} {\bibfnamefont {G.}~\bibnamefont {Shakya}}, \bibinfo {author} {\bibfnamefont {M.}~\bibnamefont {Cattaneo}}, \bibinfo {author} {\bibfnamefont {G.}~\bibnamefont {Guerriero}}, \bibinfo {author} {\bibfnamefont {A.}~\bibnamefont {Prasanna}}, \bibinfo {author} {\bibfnamefont {S.}~\bibnamefont {Fiorini}},\ and\ \bibinfo {author} {\bibfnamefont {O.}~\bibnamefont {Supponen}},\ }\href@noop {} {\bibfield  {journal} {\bibinfo  {journal} {Advanced Drug Delivery Reviews}\ }\textbf {\bibinfo {volume} {206}} (\bibinfo {year} {2024})}\BibitemShut {NoStop}%
\bibitem [{\citenamefont {Rossell{\'{o}}}\ \emph {et~al.}(2018)\citenamefont {Rossell{\'{o}}}, \citenamefont {Lauterborn}, \citenamefont {Koch}, \citenamefont {Wilken}, \citenamefont {Kurz},\ and\ \citenamefont {Mettin}}]{Rossello2018AcousticallyJets}%
  \BibitemOpen
  \bibfield  {author} {\bibinfo {author} {\bibfnamefont {J.~M.}\ \bibnamefont {Rossell{\'{o}}}}, \bibinfo {author} {\bibfnamefont {W.}~\bibnamefont {Lauterborn}}, \bibinfo {author} {\bibfnamefont {M.}~\bibnamefont {Koch}}, \bibinfo {author} {\bibfnamefont {T.}~\bibnamefont {Wilken}}, \bibinfo {author} {\bibfnamefont {T.}~\bibnamefont {Kurz}},\ and\ \bibinfo {author} {\bibfnamefont {R.}~\bibnamefont {Mettin}},\ }\href@noop {} {\bibfield  {journal} {\bibinfo  {journal} {Physics of Fluids}\ }\textbf {\bibinfo {volume} {30}} (\bibinfo {year} {2018})}\BibitemShut {NoStop}%
\bibitem [{\citenamefont {Cleve}\ \emph {et~al.}(2019)\citenamefont {Cleve}, \citenamefont {Inserra},\ and\ \citenamefont {Prentice}}]{Cleve2019ContrastUltrasound}%
  \BibitemOpen
  \bibfield  {author} {\bibinfo {author} {\bibfnamefont {S.}~\bibnamefont {Cleve}}, \bibinfo {author} {\bibfnamefont {C.}~\bibnamefont {Inserra}},\ and\ \bibinfo {author} {\bibfnamefont {P.}~\bibnamefont {Prentice}},\ }\href@noop {} {\bibfield  {journal} {\bibinfo  {journal} {Ultrasound in Medicine and Biology}\ }\textbf {\bibinfo {volume} {45}} (\bibinfo {year} {2019})}\BibitemShut {NoStop}%
\bibitem [{\citenamefont {Ohl}\ and\ \citenamefont {Ikink}(2003)}]{Ohl2003Shock-Wave-InducedBubbles}%
  \BibitemOpen
  \bibfield  {author} {\bibinfo {author} {\bibfnamefont {C.~D.}\ \bibnamefont {Ohl}}\ and\ \bibinfo {author} {\bibfnamefont {R.}~\bibnamefont {Ikink}},\ }\href@noop {} {\bibfield  {journal} {\bibinfo  {journal} {Physical Review Letters}\ }\textbf {\bibinfo {volume} {90}} (\bibinfo {year} {2003})}\BibitemShut {NoStop}%
\bibitem [{\citenamefont {Bokman}\ \emph {et~al.}(2023)\citenamefont {Bokman}, \citenamefont {Biasiori-Poulanges}, \citenamefont {Meyer},\ and\ \citenamefont {Supponen}}]{Bokman2023ScalingWave}%
  \BibitemOpen
  \bibfield  {author} {\bibinfo {author} {\bibfnamefont {G.~T.}\ \bibnamefont {Bokman}}, \bibinfo {author} {\bibfnamefont {L.}~\bibnamefont {Biasiori-Poulanges}}, \bibinfo {author} {\bibfnamefont {D.~W.}\ \bibnamefont {Meyer}},\ and\ \bibinfo {author} {\bibfnamefont {O.}~\bibnamefont {Supponen}},\ }\href@noop {} {\bibfield  {journal} {\bibinfo  {journal} {Journal of Fluid Mechanics}\ }\textbf {\bibinfo {volume} {967}} (\bibinfo {year} {2023})}\BibitemShut {NoStop}%
\bibitem [{\citenamefont {Blake}\ and\ \citenamefont {Gibson}(1987)}]{Blake1987CAVITATIONBOUNDARIES.}%
  \BibitemOpen
  \bibfield  {author} {\bibinfo {author} {\bibfnamefont {J.~R.}\ \bibnamefont {Blake}}\ and\ \bibinfo {author} {\bibfnamefont {D.~C.}\ \bibnamefont {Gibson}},\ }\href@noop {} {\bibfield  {journal} {\bibinfo  {journal} {Annual Review of Fluid Mechanics}\ }\textbf {\bibinfo {volume} {19}} (\bibinfo {year} {1987})}\BibitemShut {NoStop}%
\bibitem [{\citenamefont {Supponen}\ \emph {et~al.}(2016)\citenamefont {Supponen}, \citenamefont {Obreschkow}, \citenamefont {Tinguely}, \citenamefont {Kobel}, \citenamefont {Dorsaz},\ and\ \citenamefont {Farhat}}]{Supponen2016ScalingBubbles}%
  \BibitemOpen
  \bibfield  {author} {\bibinfo {author} {\bibfnamefont {O.}~\bibnamefont {Supponen}}, \bibinfo {author} {\bibfnamefont {D.}~\bibnamefont {Obreschkow}}, \bibinfo {author} {\bibfnamefont {M.}~\bibnamefont {Tinguely}}, \bibinfo {author} {\bibfnamefont {P.}~\bibnamefont {Kobel}}, \bibinfo {author} {\bibfnamefont {N.}~\bibnamefont {Dorsaz}},\ and\ \bibinfo {author} {\bibfnamefont {M.}~\bibnamefont {Farhat}},\ }\href@noop {} {\bibfield  {journal} {\bibinfo  {journal} {Journal of Fluid Mechanics}\ }\textbf {\bibinfo {volume} {802}},\ \bibinfo {pages} {263} (\bibinfo {year} {2016})}\BibitemShut {NoStop}%
\bibitem [{\citenamefont {Duchemin}\ \emph {et~al.}(2002)\citenamefont {Duchemin}, \citenamefont {Popinet}, \citenamefont {Josserand},\ and\ \citenamefont {Zaleski}}]{Duchemin2002JetSurface}%
  \BibitemOpen
  \bibfield  {author} {\bibinfo {author} {\bibfnamefont {L.}~\bibnamefont {Duchemin}}, \bibinfo {author} {\bibfnamefont {S.}~\bibnamefont {Popinet}}, \bibinfo {author} {\bibfnamefont {C.}~\bibnamefont {Josserand}},\ and\ \bibinfo {author} {\bibfnamefont {S.}~\bibnamefont {Zaleski}},\ }\href@noop {} {\bibfield  {journal} {\bibinfo  {journal} {Physics of Fluids}\ }\textbf {\bibinfo {volume} {14}} (\bibinfo {year} {2002})}\BibitemShut {NoStop}%
\bibitem [{\citenamefont {Ghabache}\ \emph {et~al.}(2014)\citenamefont {Ghabache}, \citenamefont {Antkowiak}, \citenamefont {Josserand},\ and\ \citenamefont {S{\'{e}}on}}]{Ghabache2014OnEjection}%
  \BibitemOpen
  \bibfield  {author} {\bibinfo {author} {\bibfnamefont {E.}~\bibnamefont {Ghabache}}, \bibinfo {author} {\bibfnamefont {A.}~\bibnamefont {Antkowiak}}, \bibinfo {author} {\bibfnamefont {C.}~\bibnamefont {Josserand}},\ and\ \bibinfo {author} {\bibfnamefont {T.}~\bibnamefont {S{\'{e}}on}},\ }\href@noop {} {\bibfield  {journal} {\bibinfo  {journal} {Physics of Fluids}\ }\textbf {\bibinfo {volume} {26}} (\bibinfo {year} {2014})}\BibitemShut {NoStop}%
\bibitem [{\citenamefont {Burton}\ \emph {et~al.}(2005)\citenamefont {Burton}, \citenamefont {Waldrep},\ and\ \citenamefont {Taborek}}]{Burton2005ScalingPinch-off}%
  \BibitemOpen
  \bibfield  {author} {\bibinfo {author} {\bibfnamefont {J.~C.}\ \bibnamefont {Burton}}, \bibinfo {author} {\bibfnamefont {R.}~\bibnamefont {Waldrep}},\ and\ \bibinfo {author} {\bibfnamefont {P.}~\bibnamefont {Taborek}},\ }\href@noop {} {\bibfield  {journal} {\bibinfo  {journal} {Physical Review Letters}\ }\textbf {\bibinfo {volume} {94}} (\bibinfo {year} {2005})}\BibitemShut {NoStop}%
\bibitem [{\citenamefont {Gordillo}\ \emph {et~al.}(2005)\citenamefont {Gordillo}, \citenamefont {Sevilla}, \citenamefont {Rodr{\'{i}}guez-Rodr{\'{i}}guez},\ and\ \citenamefont {Mart{\'{i}}nez-Baz{\'{a}}n}}]{Gordillo2005AxisymmetricNumbers}%
  \BibitemOpen
  \bibfield  {author} {\bibinfo {author} {\bibfnamefont {J.~M.}\ \bibnamefont {Gordillo}}, \bibinfo {author} {\bibfnamefont {A.}~\bibnamefont {Sevilla}}, \bibinfo {author} {\bibfnamefont {J.}~\bibnamefont {Rodr{\'{i}}guez-Rodr{\'{i}}guez}},\ and\ \bibinfo {author} {\bibfnamefont {C.}~\bibnamefont {Mart{\'{i}}nez-Baz{\'{a}}n}},\ }\href@noop {} {\bibfield  {journal} {\bibinfo  {journal} {Physical Review Letters}\ }\textbf {\bibinfo {volume} {95}} (\bibinfo {year} {2005})}\BibitemShut {NoStop}%
\bibitem [{\citenamefont {Jiang}\ \emph {et~al.}(2024)\citenamefont {Jiang}, \citenamefont {Rotily}, \citenamefont {Villermaux},\ and\ \citenamefont {Wang}}]{Jiang2024AbyssCollisions}%
  \BibitemOpen
  \bibfield  {author} {\bibinfo {author} {\bibfnamefont {X.}~\bibnamefont {Jiang}}, \bibinfo {author} {\bibfnamefont {L.}~\bibnamefont {Rotily}}, \bibinfo {author} {\bibfnamefont {E.}~\bibnamefont {Villermaux}},\ and\ \bibinfo {author} {\bibfnamefont {X.}~\bibnamefont {Wang}},\ }\href@noop {} {\bibfield  {journal} {\bibinfo  {journal} {Physical Review Letters}\ }\textbf {\bibinfo {volume} {133}} (\bibinfo {year} {2024})}\BibitemShut {NoStop}%
\bibitem [{\citenamefont {Lechner}\ \emph {et~al.}(2019)\citenamefont {Lechner}, \citenamefont {Lauterborn}, \citenamefont {Koch},\ and\ \citenamefont {Mettin}}]{Lechner2019FastStudy}%
  \BibitemOpen
  \bibfield  {author} {\bibinfo {author} {\bibfnamefont {C.}~\bibnamefont {Lechner}}, \bibinfo {author} {\bibfnamefont {W.}~\bibnamefont {Lauterborn}}, \bibinfo {author} {\bibfnamefont {M.}~\bibnamefont {Koch}},\ and\ \bibinfo {author} {\bibfnamefont {R.}~\bibnamefont {Mettin}},\ }\href@noop {} {\bibfield  {journal} {\bibinfo  {journal} {Physical Review Fluids}\ }\textbf {\bibinfo {volume} {4}} (\bibinfo {year} {2019})}\BibitemShut {NoStop}%
\bibitem [{\citenamefont {Reuter}\ and\ \citenamefont {Ohl}(2021)}]{Reuter2021SupersonicBubbles}%
  \BibitemOpen
  \bibfield  {author} {\bibinfo {author} {\bibfnamefont {F.}~\bibnamefont {Reuter}}\ and\ \bibinfo {author} {\bibfnamefont {C.~D.}\ \bibnamefont {Ohl}},\ }\href@noop {} {\bibfield  {journal} {\bibinfo  {journal} {Applied Physics Letters}\ }\textbf {\bibinfo {volume} {118}} (\bibinfo {year} {2021})}\BibitemShut {NoStop}%
\bibitem [{\citenamefont {Crum}(1979)}]{Crum1979SurfaceBubbles}%
  \BibitemOpen
  \bibfield  {author} {\bibinfo {author} {\bibfnamefont {L.~A.}\ \bibnamefont {Crum}},\ }\href@noop {} {\bibfield  {journal} {\bibinfo  {journal} {J. Phys. Colloques}\ }\textbf {\bibinfo {volume} {40}},\ \bibinfo {pages} {8} (\bibinfo {year} {1979})}\BibitemShut {NoStop}%
\bibitem [{\citenamefont {Prabowo}\ and\ \citenamefont {Ohl}(2011)}]{Prabowo2011SurfaceBubbles}%
  \BibitemOpen
  \bibfield  {author} {\bibinfo {author} {\bibfnamefont {F.}~\bibnamefont {Prabowo}}\ and\ \bibinfo {author} {\bibfnamefont {C.~D.}\ \bibnamefont {Ohl}},\ }\href@noop {} {\bibfield  {journal} {\bibinfo  {journal} {Ultrasonics Sonochemistry}\ }\textbf {\bibinfo {volume} {18}} (\bibinfo {year} {2011})}\BibitemShut {NoStop}%
\bibitem [{\citenamefont {Vos}\ \emph {et~al.}(2011)\citenamefont {Vos}, \citenamefont {Dollet}, \citenamefont {Versluis},\ and\ \citenamefont {De~Jong}}]{Vos2011}%
  \BibitemOpen
  \bibfield  {author} {\bibinfo {author} {\bibfnamefont {H.~J.}\ \bibnamefont {Vos}}, \bibinfo {author} {\bibfnamefont {B.}~\bibnamefont {Dollet}}, \bibinfo {author} {\bibfnamefont {M.}~\bibnamefont {Versluis}},\ and\ \bibinfo {author} {\bibfnamefont {N.}~\bibnamefont {De~Jong}},\ }\href@noop {} {\bibfield  {journal} {\bibinfo  {journal} {Ultrasound in Medicine and Biology}\ }\textbf {\bibinfo {volume} {37}},\ \bibinfo {pages} {935} (\bibinfo {year} {2011})}\BibitemShut {NoStop}%
\bibitem [{\citenamefont {Biasiori-Poulanges}\ \emph {et~al.}(2023)\citenamefont {Biasiori-Poulanges}, \citenamefont {Bourquard}, \citenamefont {Luki{\'{c}}}, \citenamefont {Broche},\ and\ \citenamefont {Supponen}}]{Biasiori-Poulanges2023SynchrotronCavitation}%
  \BibitemOpen
  \bibfield  {author} {\bibinfo {author} {\bibfnamefont {L.}~\bibnamefont {Biasiori-Poulanges}}, \bibinfo {author} {\bibfnamefont {C.}~\bibnamefont {Bourquard}}, \bibinfo {author} {\bibfnamefont {B.}~\bibnamefont {Luki{\'{c}}}}, \bibinfo {author} {\bibfnamefont {L.}~\bibnamefont {Broche}},\ and\ \bibinfo {author} {\bibfnamefont {O.}~\bibnamefont {Supponen}},\ }\href@noop {} {\bibfield  {journal} {\bibinfo  {journal} {Ultrasonics Sonochemistry}\ }\textbf {\bibinfo {volume} {92}} (\bibinfo {year} {2023})}\BibitemShut {NoStop}%
\bibitem [{\citenamefont {Cattaneo}\ \emph {et~al.}(2025{\natexlab{a}})\citenamefont {Cattaneo}, \citenamefont {Guerriero}, \citenamefont {Shakya}, \citenamefont {Krattiger}, \citenamefont {G.~Paganella}, \citenamefont {Narciso},\ and\ \citenamefont {Supponen}}]{Cattaneo2025CyclicDeliveryb}%
  \BibitemOpen
  \bibfield  {author} {\bibinfo {author} {\bibfnamefont {M.}~\bibnamefont {Cattaneo}}, \bibinfo {author} {\bibfnamefont {G.}~\bibnamefont {Guerriero}}, \bibinfo {author} {\bibfnamefont {G.}~\bibnamefont {Shakya}}, \bibinfo {author} {\bibfnamefont {L.~A.}\ \bibnamefont {Krattiger}}, \bibinfo {author} {\bibfnamefont {L.}~\bibnamefont {G.~Paganella}}, \bibinfo {author} {\bibfnamefont {M.~L.}\ \bibnamefont {Narciso}},\ and\ \bibinfo {author} {\bibfnamefont {O.}~\bibnamefont {Supponen}},\ }\href {https://doi.org/10.1038/s41567-025-02785-0} {\bibfield  {journal} {\bibinfo  {journal} {Nature Physics}\ }\textbf {\bibinfo {volume} {21}},\ \bibinfo {pages} {590} (\bibinfo {year} {2025}{\natexlab{a}})}\BibitemShut {NoStop}%
\bibitem [{\citenamefont {Longuet-Higgins}(1983)}]{Longuet-Higgins1983BubblesSurface}%
  \BibitemOpen
  \bibfield  {author} {\bibinfo {author} {\bibfnamefont {M.~S.}\ \bibnamefont {Longuet-Higgins}},\ }\href@noop {} {\bibfield  {journal} {\bibinfo  {journal} {Journal of Fluid Mechanics}\ }\textbf {\bibinfo {volume} {127}} (\bibinfo {year} {1983})}\BibitemShut {NoStop}%
\bibitem [{\citenamefont {Goodridge}\ \emph {et~al.}(1996)\citenamefont {Goodridge}, \citenamefont {Shi},\ and\ \citenamefont {Lathrop}}]{Goodridge1996ThresholdWaves}%
  \BibitemOpen
  \bibfield  {author} {\bibinfo {author} {\bibfnamefont {C.~L.}\ \bibnamefont {Goodridge}}, \bibinfo {author} {\bibfnamefont {W.~T.}\ \bibnamefont {Shi}},\ and\ \bibinfo {author} {\bibfnamefont {D.~P.}\ \bibnamefont {Lathrop}},\ }\href@noop {} {\bibfield  {journal} {\bibinfo  {journal} {Physical Review Letters}\ }\textbf {\bibinfo {volume} {76}} (\bibinfo {year} {1996})}\BibitemShut {NoStop}%
\bibitem [{\citenamefont {Goodridge}\ \emph {et~al.}(1997)\citenamefont {Goodridge}, \citenamefont {Tao~Shi}, \citenamefont {Hentschel},\ and\ \citenamefont {Lathrop}}]{Goodridge1997ViscousWaves}%
  \BibitemOpen
  \bibfield  {author} {\bibinfo {author} {\bibfnamefont {C.~L.}\ \bibnamefont {Goodridge}}, \bibinfo {author} {\bibfnamefont {W.}~\bibnamefont {Tao~Shi}}, \bibinfo {author} {\bibfnamefont {H.~G.}\ \bibnamefont {Hentschel}},\ and\ \bibinfo {author} {\bibfnamefont {D.~P.}\ \bibnamefont {Lathrop}},\ }\href@noop {} {\bibfield  {journal} {\bibinfo  {journal} {Physical Review E - Statistical Physics, Plasmas, Fluids, and Related Interdisciplinary Topics}\ }\textbf {\bibinfo {volume} {56}} (\bibinfo {year} {1997})}\BibitemShut {NoStop}%
\bibitem [{\citenamefont {Zeff}\ \emph {et~al.}(2000)\citenamefont {Zeff}, \citenamefont {Kleber}, \citenamefont {Fineberg},\ and\ \citenamefont {Lathrop}}]{Zeff2000SingularitySurface}%
  \BibitemOpen
  \bibfield  {author} {\bibinfo {author} {\bibfnamefont {B.}~\bibnamefont {Zeff}}, \bibinfo {author} {\bibfnamefont {B.}~\bibnamefont {Kleber}}, \bibinfo {author} {\bibfnamefont {J.}~\bibnamefont {Fineberg}},\ and\ \bibinfo {author} {\bibfnamefont {D.}~\bibnamefont {Lathrop}},\ }\href@noop {} {\bibfield  {journal} {\bibinfo  {journal} {Nature}\ }\textbf {\bibinfo {volume} {403}},\ \bibinfo {pages} {401} (\bibinfo {year} {2000})}\BibitemShut {NoStop}%
\bibitem [{\citenamefont {Cattaneo}\ \emph {et~al.}(2025{\natexlab{b}})\citenamefont {Cattaneo}, \citenamefont {Presse}, \citenamefont {Shakya}, \citenamefont {Renggli}, \citenamefont {Luki{\'{c}}}, \citenamefont {Prasanna}, \citenamefont {Meyer}, \citenamefont {Rack},\ and\ \citenamefont {Supponen}}]{Cattaneo2025ShapeBubblesb}%
  \BibitemOpen
  \bibfield  {author} {\bibinfo {author} {\bibfnamefont {M.}~\bibnamefont {Cattaneo}}, \bibinfo {author} {\bibfnamefont {L.}~\bibnamefont {Presse}}, \bibinfo {author} {\bibfnamefont {G.}~\bibnamefont {Shakya}}, \bibinfo {author} {\bibfnamefont {T.}~\bibnamefont {Renggli}}, \bibinfo {author} {\bibfnamefont {B.}~\bibnamefont {Luki{\'{c}}}}, \bibinfo {author} {\bibfnamefont {A.}~\bibnamefont {Prasanna}}, \bibinfo {author} {\bibfnamefont {D.~W.}\ \bibnamefont {Meyer}}, \bibinfo {author} {\bibfnamefont {A.}~\bibnamefont {Rack}},\ and\ \bibinfo {author} {\bibfnamefont {O.}~\bibnamefont {Supponen}},\ }\href {https://doi.org/DOI: 10.1017/jfm.2025.10457} {\bibfield  {journal} {\bibinfo  {journal} {Journal of Fluid Mechanics}\ }\textbf {\bibinfo {volume} {1017}},\ \bibinfo {pages} {A9} (\bibinfo {year} {2025}{\natexlab{b}})}\BibitemShut {NoStop}%
\bibitem [{\citenamefont {Gekle}\ \emph {et~al.}(2009)\citenamefont {Gekle}, \citenamefont {Gordillo}, \citenamefont {Van Der~Meer},\ and\ \citenamefont {Lohse}}]{Gekle2009High-speedImpact}%
  \BibitemOpen
  \bibfield  {author} {\bibinfo {author} {\bibfnamefont {S.}~\bibnamefont {Gekle}}, \bibinfo {author} {\bibfnamefont {J.~M.}\ \bibnamefont {Gordillo}}, \bibinfo {author} {\bibfnamefont {D.}~\bibnamefont {Van Der~Meer}},\ and\ \bibinfo {author} {\bibfnamefont {D.}~\bibnamefont {Lohse}},\ }\href@noop {} {\bibfield  {journal} {\bibinfo  {journal} {Physical Review Letters}\ }\textbf {\bibinfo {volume} {102}} (\bibinfo {year} {2009})}\BibitemShut {NoStop}%
\bibitem [{\citenamefont {Gekle}\ and\ \citenamefont {Gordillo}(2010)}]{Gekle2010GenerationFormation}%
  \BibitemOpen
  \bibfield  {author} {\bibinfo {author} {\bibfnamefont {S.}~\bibnamefont {Gekle}}\ and\ \bibinfo {author} {\bibfnamefont {J.~M.}\ \bibnamefont {Gordillo}},\ }\href@noop {} {\bibfield  {journal} {\bibinfo  {journal} {Journal of Fluid Mechanics}\ }\textbf {\bibinfo {volume} {663}} (\bibinfo {year} {2010})}\BibitemShut {NoStop}%
\bibitem [{\citenamefont {Keller}\ and\ \citenamefont {Miksis}(1983)}]{Keller1983SurfaceFlows}%
  \BibitemOpen
  \bibfield  {author} {\bibinfo {author} {\bibfnamefont {J.}~\bibnamefont {Keller}}\ and\ \bibinfo {author} {\bibfnamefont {M.~J.}\ \bibnamefont {Miksis}},\ }\href@noop {} {\bibfield  {journal} {\bibinfo  {journal} {SIAM Journal on Applied Mathematics}\ }\textbf {\bibinfo {volume} {43}} (\bibinfo {year} {1983})}\BibitemShut {NoStop}%
\bibitem [{See()}]{SeeScaling.}%
  \BibitemOpen
  \href@noop {} {\bibinfo {title} {{See Supplemental Material at http://link.aps.org/supplemental/10.1103/7c4p-y9jr for details on the derivation of the self-similar scaling.}}}\BibitemShut {Stop}%
\bibitem [{\citenamefont {Lai}\ \emph {et~al.}(2018)\citenamefont {Lai}, \citenamefont {Eggers},\ and\ \citenamefont {Deike}}]{Lai2018BubbleProfiles}%
  \BibitemOpen
  \bibfield  {author} {\bibinfo {author} {\bibfnamefont {C.~Y.}\ \bibnamefont {Lai}}, \bibinfo {author} {\bibfnamefont {J.}~\bibnamefont {Eggers}},\ and\ \bibinfo {author} {\bibfnamefont {L.}~\bibnamefont {Deike}},\ }\href@noop {} {\bibfield  {journal} {\bibinfo  {journal} {Physical Review Letters}\ }\textbf {\bibinfo {volume} {121}} (\bibinfo {year} {2018})}\BibitemShut {NoStop}%
\bibitem [{\citenamefont {Bu{\ss}mann}\ \emph {et~al.}(2023)\citenamefont {Bu{\ss}mann}, \citenamefont {Riahi}, \citenamefont {G{\"{o}}kce}, \citenamefont {Adami}, \citenamefont {Barcikowski},\ and\ \citenamefont {Adams}}]{Bumann2023InvestigationSimulation}%
  \BibitemOpen
  \bibfield  {author} {\bibinfo {author} {\bibfnamefont {A.}~\bibnamefont {Bu{\ss}mann}}, \bibinfo {author} {\bibfnamefont {F.}~\bibnamefont {Riahi}}, \bibinfo {author} {\bibfnamefont {B.}~\bibnamefont {G{\"{o}}kce}}, \bibinfo {author} {\bibfnamefont {S.}~\bibnamefont {Adami}}, \bibinfo {author} {\bibfnamefont {S.}~\bibnamefont {Barcikowski}},\ and\ \bibinfo {author} {\bibfnamefont {N.~A.}\ \bibnamefont {Adams}},\ }\href@noop {} {\bibfield  {journal} {\bibinfo  {journal} {Physics of Fluids}\ }\textbf {\bibinfo {volume} {35}} (\bibinfo {year} {2023})}\BibitemShut {NoStop}%
\bibitem [{\citenamefont {Sieber}\ \emph {et~al.}(2023)\citenamefont {Sieber}, \citenamefont {Preso},\ and\ \citenamefont {Farhat}}]{Sieber2023CavitationMaterials}%
  \BibitemOpen
  \bibfield  {author} {\bibinfo {author} {\bibfnamefont {A.~B.}\ \bibnamefont {Sieber}}, \bibinfo {author} {\bibfnamefont {D.~B.}\ \bibnamefont {Preso}},\ and\ \bibinfo {author} {\bibfnamefont {M.}~\bibnamefont {Farhat}},\ }\href@noop {} {\bibfield  {journal} {\bibinfo  {journal} {Physics of Fluids}\ }\textbf {\bibinfo {volume} {35}} (\bibinfo {year} {2023})}\BibitemShut {NoStop}%
\bibitem [{\citenamefont {Leppinen}\ and\ \citenamefont {Lister}(2003)}]{Leppinen2003CapillaryFluids}%
  \BibitemOpen
  \bibfield  {author} {\bibinfo {author} {\bibfnamefont {D.}~\bibnamefont {Leppinen}}\ and\ \bibinfo {author} {\bibfnamefont {J.~R.}\ \bibnamefont {Lister}},\ }\href@noop {} {\bibfield  {journal} {\bibinfo  {journal} {Physics of Fluids}\ }\textbf {\bibinfo {volume} {15}} (\bibinfo {year} {2003})}\BibitemShut {NoStop}%
\bibitem [{\citenamefont {Thoroddsen}\ \emph {et~al.}(2007)\citenamefont {Thoroddsen}, \citenamefont {Etoh},\ and\ \citenamefont {Takehara}}]{Thoroddsen2007ExperimentsPinch-off}%
  \BibitemOpen
  \bibfield  {author} {\bibinfo {author} {\bibfnamefont {S.~T.}\ \bibnamefont {Thoroddsen}}, \bibinfo {author} {\bibfnamefont {T.~G.}\ \bibnamefont {Etoh}},\ and\ \bibinfo {author} {\bibfnamefont {K.}~\bibnamefont {Takehara}},\ }\href@noop {} {\bibfield  {journal} {\bibinfo  {journal} {Physics of Fluids}\ }\textbf {\bibinfo {volume} {19}} (\bibinfo {year} {2007})}\BibitemShut {NoStop}%
\bibitem [{\citenamefont {Eggers}\ \emph {et~al.}(2007)\citenamefont {Eggers}, \citenamefont {Fontelos}, \citenamefont {Leppinen},\ and\ \citenamefont {Snoeijer}}]{Eggers2007TheoryCavity}%
  \BibitemOpen
  \bibfield  {author} {\bibinfo {author} {\bibfnamefont {J.}~\bibnamefont {Eggers}}, \bibinfo {author} {\bibfnamefont {M.~A.}\ \bibnamefont {Fontelos}}, \bibinfo {author} {\bibfnamefont {D.}~\bibnamefont {Leppinen}},\ and\ \bibinfo {author} {\bibfnamefont {J.~H.}\ \bibnamefont {Snoeijer}},\ }\href@noop {} {\bibfield  {journal} {\bibinfo  {journal} {Physical Review Letters}\ }\textbf {\bibinfo {volume} {98}} (\bibinfo {year} {2007})}\BibitemShut {NoStop}%
\bibitem [{\citenamefont {Gordillo}\ and\ \citenamefont {Blanco-Rodr{\'{i}}guez}(2023)}]{Gordillo2023TheoryJets}%
  \BibitemOpen
  \bibfield  {author} {\bibinfo {author} {\bibfnamefont {J.~M.}\ \bibnamefont {Gordillo}}\ and\ \bibinfo {author} {\bibfnamefont {F.~J.}\ \bibnamefont {Blanco-Rodr{\'{i}}guez}},\ }\href@noop {} {\bibfield  {journal} {\bibinfo  {journal} {Physical Review Fluids}\ }\textbf {\bibinfo {volume} {8}} (\bibinfo {year} {2023})}\BibitemShut {NoStop}%
\bibitem [{\citenamefont {Michell}(1893)}]{Michell1893Water}%
  \BibitemOpen
  \bibfield  {author} {\bibinfo {author} {\bibfnamefont {J.~H.}\ \bibnamefont {Michell}},\ }\href@noop {} {\bibfield  {journal} {\bibinfo  {journal} {The London, Edinburgh, and Dublin Philosophical Magazine and Journal of Science}\ }\textbf {\bibinfo {volume} {36}} (\bibinfo {year} {1893})}\BibitemShut {NoStop}%
\bibitem [{\citenamefont {Schwartz}\ and\ \citenamefont {Vanden-Broeck}(1979)}]{Schwartz1979NumericalWaves}%
  \BibitemOpen
  \bibfield  {author} {\bibinfo {author} {\bibfnamefont {L.~W.}\ \bibnamefont {Schwartz}}\ and\ \bibinfo {author} {\bibfnamefont {J.~M.}\ \bibnamefont {Vanden-Broeck}},\ }\href@noop {} {\bibfield  {journal} {\bibinfo  {journal} {Journal of Fluid Mechanics}\ }\textbf {\bibinfo {volume} {95}} (\bibinfo {year} {1979})}\BibitemShut {NoStop}%
\bibitem [{\citenamefont {Hogrefe}\ \emph {et~al.}(1998)\citenamefont {Hogrefe}, \citenamefont {Peffley}, \citenamefont {Goodridge}, \citenamefont {Shi}, \citenamefont {Hentschel},\ and\ \citenamefont {Lathrop}}]{Hogrefe1998Power-lawWaves}%
  \BibitemOpen
  \bibfield  {author} {\bibinfo {author} {\bibfnamefont {J.~E.}\ \bibnamefont {Hogrefe}}, \bibinfo {author} {\bibfnamefont {N.~L.}\ \bibnamefont {Peffley}}, \bibinfo {author} {\bibfnamefont {C.~L.}\ \bibnamefont {Goodridge}}, \bibinfo {author} {\bibfnamefont {W.~T.}\ \bibnamefont {Shi}}, \bibinfo {author} {\bibfnamefont {H.~G.~E.}\ \bibnamefont {Hentschel}},\ and\ \bibinfo {author} {\bibfnamefont {D.~P.}\ \bibnamefont {Lathrop}},\ }\href@noop {} {\bibfield  {journal} {\bibinfo  {journal} {Physica D: Nonlinear Phenomena}\ }\textbf {\bibinfo {volume} {123}} (\bibinfo {year} {1998})}\BibitemShut {NoStop}%
\bibitem [{\citenamefont {Deike}\ \emph {et~al.}(2018)\citenamefont {Deike}, \citenamefont {Ghabache}, \citenamefont {Liger-Belair}, \citenamefont {Das}, \citenamefont {Zaleski}, \citenamefont {Popinet},\ and\ \citenamefont {S{\'{e}}on}}]{Deike2018DynamicsBubbles}%
  \BibitemOpen
  \bibfield  {author} {\bibinfo {author} {\bibfnamefont {L.}~\bibnamefont {Deike}}, \bibinfo {author} {\bibfnamefont {E.}~\bibnamefont {Ghabache}}, \bibinfo {author} {\bibfnamefont {G.}~\bibnamefont {Liger-Belair}}, \bibinfo {author} {\bibfnamefont {A.~K.}\ \bibnamefont {Das}}, \bibinfo {author} {\bibfnamefont {S.}~\bibnamefont {Zaleski}}, \bibinfo {author} {\bibfnamefont {S.}~\bibnamefont {Popinet}},\ and\ \bibinfo {author} {\bibfnamefont {T.}~\bibnamefont {S{\'{e}}on}},\ }\href@noop {} {\bibfield  {journal} {\bibinfo  {journal} {Physical Review Fluids}\ }\textbf {\bibinfo {volume} {3}} (\bibinfo {year} {2018})}\BibitemShut {NoStop}%
\bibitem [{\citenamefont {Cattaneo}\ \emph {et~al.}(2025{\natexlab{c}})\citenamefont {Cattaneo}, \citenamefont {Presse},\ and\ \citenamefont {Supponen}}]{Cattaneo2025SourceHttps://doi.org/10.5281/zenodo.15085879}%
  \BibitemOpen
  \bibfield  {author} {\bibinfo {author} {\bibfnamefont {M.}~\bibnamefont {Cattaneo}}, \bibinfo {author} {\bibfnamefont {L.}~\bibnamefont {Presse}},\ and\ \bibinfo {author} {\bibfnamefont {O.}~\bibnamefont {Supponen}},\ }\href {https://doi.org/https://doi.org/10.5281/zenodo.15085879} {\bibinfo {title} {{Source data for "Faraday wave singularities trigger microbubble jetting" (1.0.0) [Data set]. Zenodo. https://doi.org/10.5281/zenodo.15085879}}} (\bibinfo {year} {2025}{\natexlab{c}})\BibitemShut {NoStop}%
\end{thebibliography}%

\end{document}